\documentclass[floats,floatfix,nofootinbib,%
superscriptaddress]{revtex4}
\usepackage{amsmath}
\usepackage{amssymb}
\usepackage{amsfonts}
\usepackage[dvips]{graphicx}

\begin{document}

\title{LINEAR RESPONSE IN INFINITE NUCLEAR MATTER AS\\
A TOOL TO REVEAL FINITE SIZE INSTABILITIES\footnote{
Based on talk presented at 18th Nuclear Physics Workshop "Maria and Pierre
Curie", 2011, Kazimierz, Poland.
}}

\author{A. Pastore}
\affiliation{Universit\'e de Lyon, Universit\'e Lyon 1, CNRS/IN2P3\\
Institut de Physique Nucl\'eaire de Lyon,\\
F-69622 Villeurbanne cedex, France}

\author{K. Bennaceur}
\affiliation{Universit\'e de Lyon, Universit\'e Lyon 1, CNRS/IN2P3\\
Institut de Physique Nucl\'eaire de Lyon,\\
F-69622 Villeurbanne cedex, France}

\author{D. Davesne}
\affiliation{Universit\'e de Lyon, Universit\'e Lyon 1, CNRS/IN2P3\\
Institut de Physique Nucl\'eaire de Lyon,\\
F-69622 Villeurbanne cedex, France}

\author{J. Meyer}
\affiliation{Universit\'e de Lyon, Universit\'e Lyon 1, CNRS/IN2P3\\
Institut de Physique Nucl\'eaire de Lyon,\\
F-69622 Villeurbanne cedex, France}


\begin{abstract}
Nuclear effective interactions are often modelled by simple
analytical expressions such as the Skyrme zero-range force.
This effective interaction depends
on a limited number of parameters that are usually fitted using
experimental data obtained from doubly magic nuclei.
It was recently shown that many Skyrme functionals
lead to the appearance of instabilities, in particular when symmetries are
broken, for example unphysical polarization of odd-even or rotating
nuclei.  In this article, we show how the formalism of the linear
response in infinite nuclear matter can be used to predict and
avoid the regions of parameters that are responsible for these
unphysical instabilities.
\end{abstract}

\maketitle

\section{Introduction}

Methods based on the mean-field approach represent the tool of choice
for the study of all atomic nuclei but the lightest. A fairly
good approximation of the mass and size of doubly magic heavy nuclei
can be obtained at the mean-field level, often called the Hartree-Fock (HF)
approximation, while the properties of other nuclei can be obtained
by using a hierarchy of refinements from the inclusion
of pairing correlations to the description of the fluctuations of collective
degrees of freedom using the Generator Coordinate Method (GCM) and projection
techniques (for symmetry restoration).

The key object to build is the Energy Density Functional (EDF). Usually, its
expression is derived from an effective interaction
like the Gogny\cite{gogny75,gogny80} or
Skyrme\cite{skyrme56,skyrme58} ones that depend on a small number of
parameters. The Skyrme interaction being of zero-range type, the obtained
EDF depends on local densities only when
the usual Slater approximation is adopted for the Coulomb contribution.
The calculation of the average value of the Skyrme interaction
in a vacuum that is a product of occupied quasiparticle states generates
all the terms of the functional in a consistent way, {\em i.e.} the
contributions of the normal and anomalous (pairing) densities as well
as the contributions of the time-even and time-odd densities.

On the one hand,
the effective character of this approach justifies the omission of certain
terms in the functional (as, for example, the so called ``${\mathbf J}^2$''
terms) or the use of different parameters for its normal and pairing
parts. Moreover, since most of the time, the procedures used to adjust the
parameters of the Skyrme interaction use only informations from time-even
systems, one can also neglect several time-odd terms in the functional. When
such a choice is done, the backward link between the functional and an
effective interaction is broken. On the other hand, it was recently realized
that the use of a functional not exactly derived from an effective interaction
leads to spurious contributions in the energy in calculations beyond
the mean-field approximation. These spurious contributions can be
removed using a procedure recently formulated.\cite{lacroix09,bender09}
It was also shown that if the effective interaction does depend on 
non integer powers of the density, other spurious contributions
pollutes the energy in beyond mean-field calculations and the aforementioned
correction procedure can not be used.\cite{duguet09}

The use of a single effective interaction with only integer powers
of the density (or with explicit three and four body interactions)
to generate all the terms of the functional would
be the ideal choice to avoid or to be able to handle correctly the
pathologies in beyond mean-field calculations. Such interactions do
not exist at this moment and their developments are only at their
early stages.\cite{washiyama,sadoudi}

The present work is devoted to the study of finite size spin and
isospin instabilities in nuclei that appear at the mean-field
level. We use a set of effective interactions that are built
following the same protocol as the one used for the famous SLy
interactions.\cite{chabanat} These interactions are supplemented by an
effective two-body zero range tensor term. Since only doubly magic nuclei and
infinite nuclear matter in its normal phase are calculated,
the choice for the interaction in the pairing channel is not in
question here. The parts of the functional that depend on the time-odd
densities are obtained from the effective interaction and are all kept
in the calculations.
Having a non-integer power of the isoscalar density $\rho_0^\alpha$ in the
density dependent term is not a problem since we do not perform
beyond-mean-field calculations here.

This article is organized as follows. In section~\ref{sec:pp} we present
and discuss the form of the Skyrme interaction used in this study
and its parameters fitting procedure is recalled in section~\ref{sec:fit}.
Examples of finite size instabilities are shown in section~\ref{sec:inst}.
Finally, in section~\ref{sec:lr}, the linear response formalism is briefly
exposed and its ability to predict and avoid the regions in
the parameters space that lead to instabilities is demonstrated.

\section{The Skyrme effective interaction}
\label{sec:pp}

The interactions and functionals used in this article to illustrate the
appearance of finite size instabilities are: (i) a serie of functionals
obtained by modifying only one coefficient on the functional derived from the
SLy5 interaction\cite{chabanat} and leaving the other coefficients unchanged;
(ii) the serie of T$IJ$ interactions
built for the study of the effect of tensor couplings on the properties
of even-even nuclei.\cite{lesinski07,bender09b}

In the T$IJ$ intaractions, the central and spin-orbit parts are of the same
form as for the SLy interactions, the tensor part being
\begin{align}
v_T(\mathbf r)&=\tfrac{1}{2}\,t_e\left\{
\delta(\mathbf r)
\left[3
\left(\boldsymbol\sigma_1\cdot\mathbf k\right)
\left(\boldsymbol\sigma_2\cdot\mathbf k\right)
-\left(\boldsymbol\sigma_1\cdot\boldsymbol\sigma_2\right)\mathbf k^2
\right]
+\mathrm{h.c.}\right\} \nonumber \\
&+\tfrac{1}{2}\,t_o\left\{
\left[3
\left(\boldsymbol\sigma_1\cdot\mathbf k'\right)\delta(\mathbf r)
\left(\boldsymbol\sigma_2\cdot\mathbf k\right)
-\left(\boldsymbol\sigma_1\cdot\boldsymbol\sigma_2\right)\mathbf k'
\cdot\delta(\mathbf r)\mathbf k
\right]
+\mathrm{h.c.}\right\}\,,
\label{vtens}
\end{align}
where h.c. stands for ``hermitian conjugate''. The parameters $t_e$
and $t_o$, along with $t_1$, $x_1$, $t_2$ and $x_2$
determine the eight coupling constants $C_t^T$, $C_t^F$, $C_t^{\Delta s}$
and $C_t^{\nabla s}$ (with $t=0$ and $1$ for the isoscalar and isovector
parts of the functional). The definitions of the operators $\mathbf k$,
$\mathbf k'$, $\boldsymbol\sigma_1$ and $\boldsymbol\sigma_2$ as well as
the list and expressions of the coupling constants appearing in the
functional can be found in the article of Lesinski
\emph{et al.}\cite{lesinski07}

\section{Parameters fitting procedure}
\label{sec:fit}

The method used to fit the parameters of a Skyrme interaction or functional
consists in constructing a merit function that measures the deviation between
targeted and calculated quantities. These quantities are generally
several properties of the infinite nuclear matter, masses and charge
radii of a set of doubly magic nuclei and one splitting between two
single particle states, this latter being added to optimize the
spin-orbit strength.

For the serie of functionals built from the SLy5 interaction, we have modified
the coefficient $C_1^{\Delta\rho}$ given by
\begin{equation}
C_1^{\Delta\rho}=\tfrac{3}{32}\,t_1\left(x_1+\tfrac{1}{2}\right)
+\tfrac{1}{32}\,t_2\left(x_2+\tfrac{1}{2}\right)\,.
\end{equation}
With the parameters of SLy5 the value of this coefficent is
$16.375~\mathrm{MeV}\,\mathrm{fm}^5$. The serie of functional used here
is then obtained by varying this coefficient without modifying the other ones.
Changing a single coupling constant without refitting the entire functional
deteriorates the properties of the functional itself (\emph{i.e.} the
reproduction of some physical observables), but this is not crucial in the
present work since this modification is only used to explore the relation
between the appearance of instabilities and the value of one term of the
functional.

The set of T$IJ$ interactions are obtained by setting
$C_0^T=30(J+I-4)~\mathrm{MeV}\,\mathrm{fm}^5$ and
$C_1^T=30(J-I)~\mathrm{MeV}\,\mathrm{fm}^5$ for $I$ and $J$ integers from
1 to 6 and with
\begin{align}
C_0^T&= -\tfrac{1}{8}\,t_1\left(x_1+\tfrac{1}{2}\right)
+\tfrac{1}{8}\,t_2\left(x_1+\tfrac{1}{2}\right)
-\tfrac{1}{8}\left(t_e+3t_o\right)\,,\\
C_1^T&= -\tfrac{1}{16}\,t_1+\tfrac{1}{16}\,t_2
-\tfrac{1}{8}\left(t_e-t_o\right)\,.
\end{align}
All the parameters of the interactions being then adjusted in order
to minimize a given merit function\cite{lesinski07} with an
iterative procedure nearly identical to that used for the
construction of the SLy parameterizations.\cite{chabanat}

\section{Instabilities in finite nuclei}
\label{sec:inst}

To adjust the parameters of an effective interaction,
one starts from an initial set of parameters that can at least provide
with a saturation point in symmetric infinite nuclear matter and
this set of parameters is modified iteratively in order to minimize
a merit function. During the iterations, some instabilities can be experienced.
For example, one can observe domains in finite nuclei where the isovector
density becomes large. This phenomenon can ultimately lead to a total
separation between protons and neutrons and a disappearance of nuclei as bound
states. The regions of the parameters space that lead to these unwanted
situations should be avoided and it would be helpful to have a criterium that
predicts where those regions are.

In a previous work,\cite{lesinski06} it was shown that the coefficient
$C_1^{\Delta\rho}$ of the Skyrme functional can be related with this kind of
isospin instability. We illustrate this correspondance by the following
schematic serie of calculations for $^{48}$Ca. We start from the effective
interaction SLy5 which is stable in the isospin channel and for which 
$C_1^{\Delta\rho}=16.375~\mathrm{MeV}\,\mathrm{fm}^5$. We keep all the other
coefficients of the functional constant and vary $C_1^{\Delta\rho}$
from $16~\mathrm{MeV}\,\mathrm{fm}^5$ to $37.1~\mathrm{MeV}\,\mathrm{fm}^5$
where an isospin instability appears in this nucleus (the limit value
can change from one nucleus to an other).

We observe that the total energy of $^{48}$Ca
calculated at the HF approximation decreases with $C_1^{\Delta\rho}$
(from -415.9~MeV to -417.2~MeV) and that this gain of energy is due to the
appearance of strong oscillations of the isovector density
$\rho_1=\rho_n-\rho_p$ as shown on Fig.~\ref{fig:iso}. For
$C_1^{\Delta\rho}\gtrsim 37.1~\mathrm{MeV}\,\mathrm{fm}^5$
the nucleus $^{48}$Ca is disrupted in finite size domains of neutrons and
protons.

This kind of instabilities can be detected during the fit procedure but
the situation is obviously more complicated for the instabilities
in the spin channel since one has to break the time reversal symmetry
to observe it. A possible constraint on the Landau parameters
(as it was the case for the SLy functionals) would only avoid
global polarizations of infinite nuclear matter,\cite{colo10} but not the
finite size instabilities.

Unfortunately, these spin instabilities have been identified recently
in mean-field calculations where the time reversal symmetry is broken
(cranked calculation\cite{hellemans} and calculations of odd-even
nuclei\cite{schunck}) when the functional is explicitely derived
from the effective interaction ({\em i.e.} all time-even and time-odd
terms are kept). Calculations of systems where the time
reversal symmetry is broken are two much time consuming to be
easily incorporated in the fit procedure of the parameters, one
needs an efficient tool to predict and avoid all possible finite
size instabilities both in the spin and isospin channels.

\section{Linear response}
\label{sec:lr}

\subsection{Formalism}

The basic formalism of linear response theory has been already explained in
several articles, in our work we follow the formalism used in the article of
Garcia-Recio \emph{et al.}\cite{garciarecio} In this work, we restrict our
study to the response to an external probe of an infinite system initially
unpolarized
both in spin and isospin spaces (symmetric nuclear matter, SNM) at zero
temperature. The excitations of the system result from the residual
particle-hole interaction between particles below and above the Fermi level.
In the theory of Fermi liquids, this interaction is given by the second
functional derivative of the Energy Functional $E[\rho]$, with respect to the
densities taken from the Hartree-Fock solution. Its matrix elements read
\begin{equation}\label{eq:matr:el}
V_{\text{ph}}^{(\alpha,\alpha')}(q,\mathbf{k}_{1},\mathbf{k}_{2})\equiv
\frac{\delta^2 E[\rho]}{\delta \rho(\mathbf{k}_1+\mathbf{q},
 \mathbf{k}_1)\delta \rho(\mathbf{k}_2, \mathbf{k}_2+\mathbf{q})} \,,
\end{equation}
where we  denote by $\alpha=\text{(S,M;I)}$ the spin and isospin in the
particle-hole (p-h) channels with $S=0$ (or 1) for the non spin-flip
(or spin-flip) channel, $I=0$ (or 1) for the isoscalar (or isovector) channel,
and $M$ being the projection of the spin on the quantization axis.
As usual, we have chosen as independent variables
the initial (final) momentum $\mathbf{k}_{1}$ ($\mathbf{k}_{2}$) of the hole
and the external momentum transfer $\mathbf{q}$. To calculate the response of
the medium, it is convenient to introduce the Green's function, or p-h
propagator, at the energy $\omega$:
$G^{(\alpha)}(\mathbf{q},\omega,\mathbf{k}_{1})$, which in the
Hartree-Fock approximation, does not depend on the spin-isospin channel
($\alpha$) and reads\cite{fetter}
\begin{equation}
G_{\text{HF}}(q,\omega,\mathbf{k}_{1})\equiv \frac{\theta(k_{F}-k_{1})
-\theta(k_{F}-|\mathbf{k}_{1}+\mathbf{q}|)}{\omega+\varepsilon(k_{1})
-\varepsilon(|\mathbf{k}_{1}+\mathbf{q}|)+i\eta \omega}\,.
\end{equation}
Considering now the residual interaction of equation~(\ref{eq:matr:el}), we can
calculate the RPA correlated Green's function by solving the Bethe-Salpeter
equation
\begin{eqnarray}
G_{\text{RPA}}^{(\alpha)}(q,\omega,\mathbf{k}_{1})
   &=&G_{\text{HF}}(q,\omega,\mathbf{k}_{1})\nonumber\\
   &+&G_{\text{HF}}(q,\omega,\mathbf{k}_{1})\sum_{\alpha'}
      \!\int\!\frac{\mathrm d^{3}k_{2}}{(2\pi)^{3}}\,
            V_{\text{ph}}^{(\alpha,\alpha')}
      (q,\mathbf{k}_{1},\mathbf{k}_{2})G_{\text{RPA}}^{(\alpha')}
      (q,\omega,\mathbf{k}_{2})\,.
\label{BSeq}
\end{eqnarray}
For a detailed discussion on the solution of equation~(\ref{BSeq}) we refer
to the work of Davesne \emph{et al.}\cite{davesne}
Finally the response function $\chi^{(\alpha)}_{\text{RPA}}(q,\omega)$ of
the infinite medium is 
\begin{equation}
\chi^{(\alpha)}_{\text{RPA}}(q,\omega)=g\int
\frac{\mathrm d^{3}k_{1}}{(2\pi)^{3}}
G_{\text{RPA}}^{(\alpha)}(q,\omega,\mathbf{k}_{1})\,,
\label{eq:chirpa}
\end{equation}
where $g$ is the spin-isospin degeneracy factor ($g=4$ for SNM).
Within this formalism, an unstable mode is marked by the occurrence of a
pole in $\chi^{(\alpha)}_{\text{RPA}}$ at $\omega=0$, corresponding to
zero excitation energy. The occurrence of a pole at finite $q$ characterizes
a system that is unstable within respect to the appearance of a spatial
oscillation of a given type with a given wavelength $\lambda\sim 2\pi/q$.

\subsection{Link with finite size instabilities in nuclei}

\begin{figure}[htbp]
$\begin{matrix}
\includegraphics[angle=270,width=6.15cm]{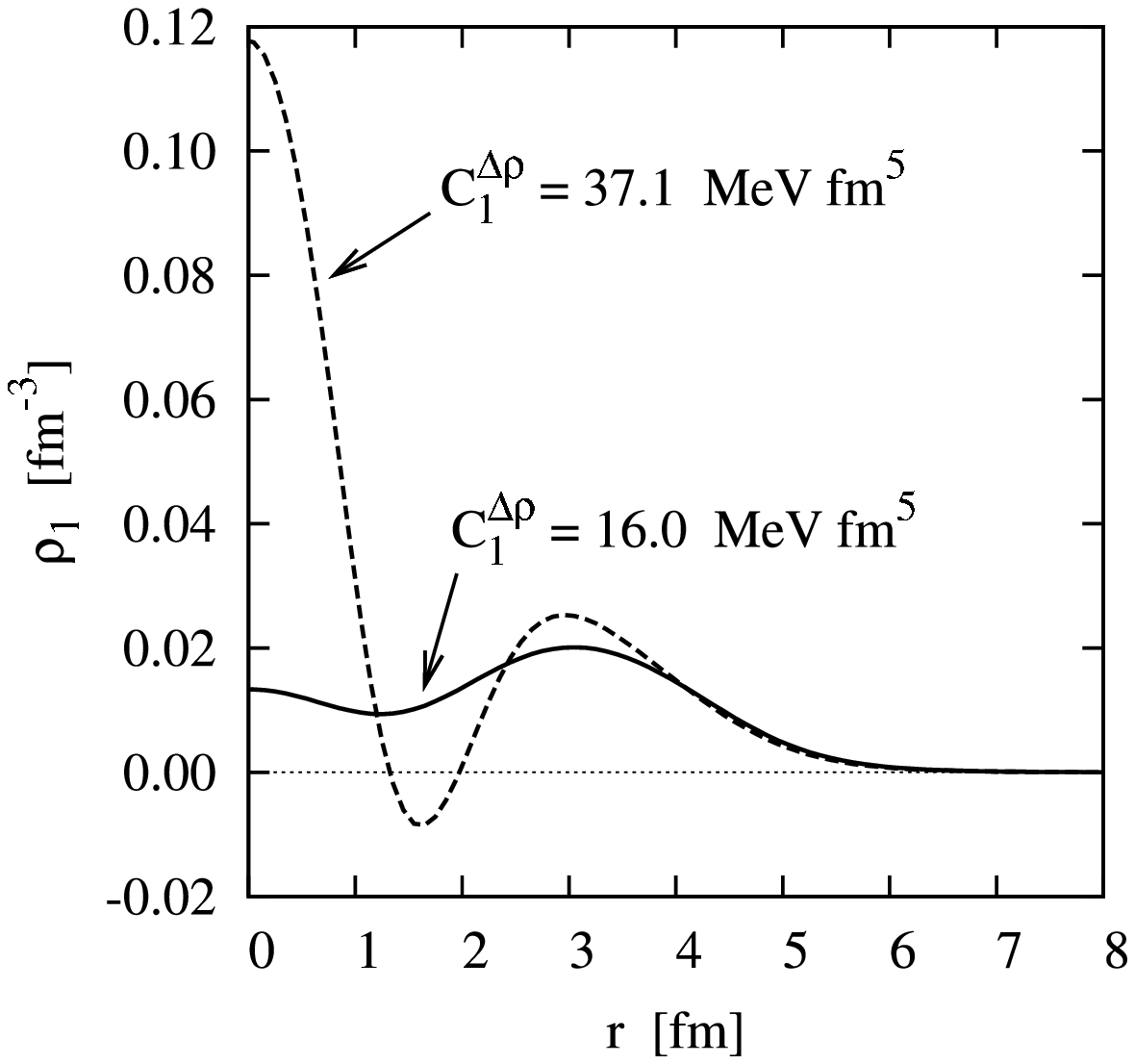} \,&\,
\includegraphics[angle=270,width=6.15cm]{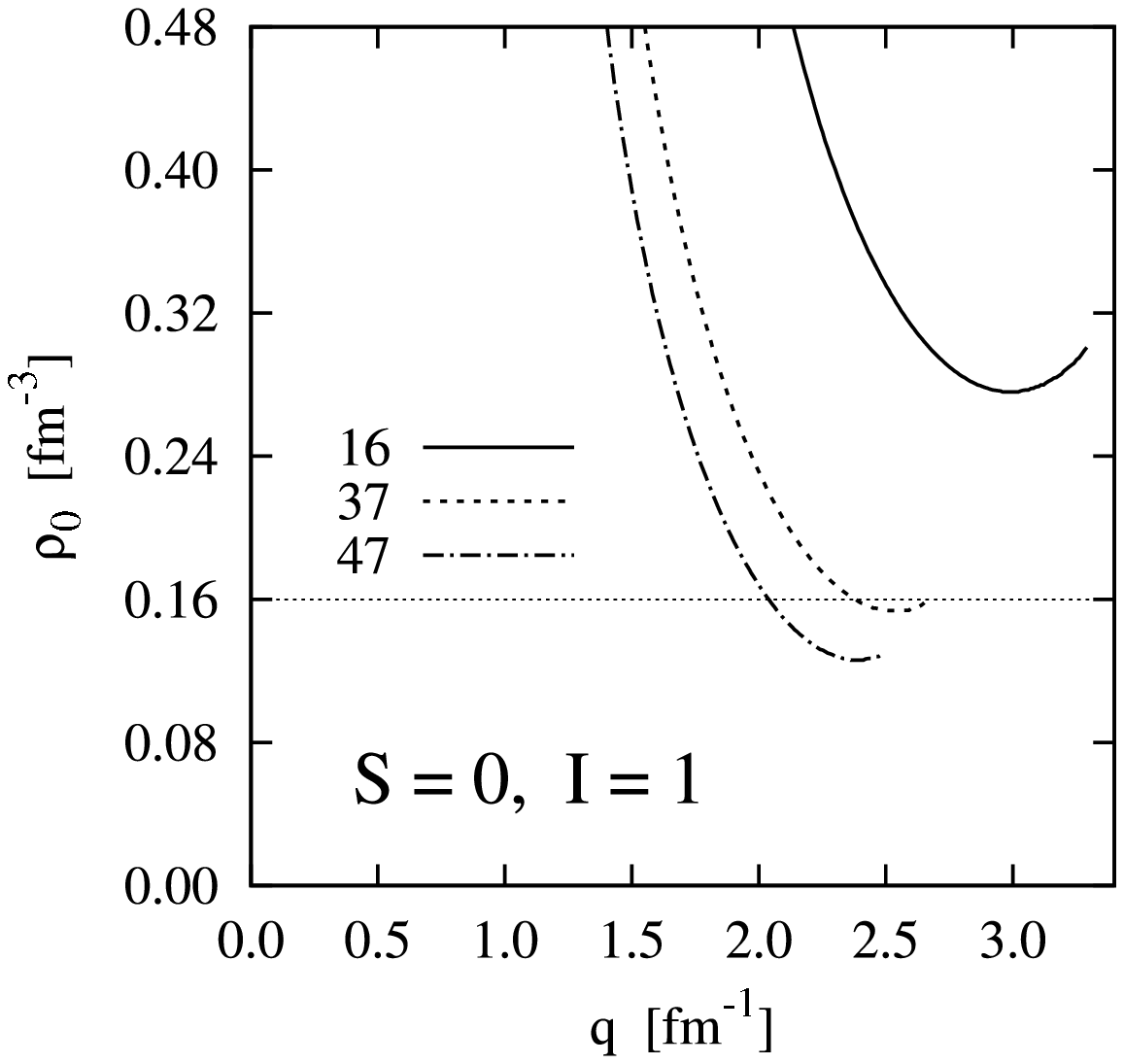}
\end{matrix}$
\vspace*{8pt}
\caption{Left panel: Isovector density $\rho_1$ for
$C_1^{\Delta\rho}=16~\mathrm{MeV}\,\mathrm{fm}^5$
and $C_1^{\Delta\rho}=37.1~\mathrm{MeV}\,\mathrm{fm}^5$ in $^{48}$Ca.
Right panel: Critical densities for three versions of the modified SLy5
functionals ($C_1^{\Delta\rho}=16$, $37$ and $47~\mathrm{MeV}\,\mathrm{fm}^5$)
in the $S=0,\,I=1$ channel. The horizontal dotted line represents the
saturation density of the system $\rho_{\text{sat}}$.
\label{fig:iso}}
\end{figure}

We search for the density $\rho_\mathrm{crit}$ in the $(q,\rho_0)$
plan that corresponds to a pole in equation~(\ref{eq:chirpa})
at zero excitation energy. These densities are represented on the
right panel of
Fig.~\ref{fig:iso} for three modified SLy5 functionals in the $S=0,\,I=1$
channels.

One can see that there is a clear relation between the instabilities
discussed in section~\ref{sec:inst} and the critical densities represented
on the right panel of Fig.~\ref{fig:iso}, in particular we observe that
the poles of the response function in the scalar-isoscalar channel approach
the saturation density $\rho_{\text{sat}}=0.16~\text{ fm}^{-3}$ for increasing
values of $C^{\Delta \rho}_1$. It is important to remark that this particular
instability can not be detected using the Landau parameters since they only
probe the limit $q\rightarrow 0$.

\begin{figure}[htbp]
$\begin{matrix}
\includegraphics[angle=270,width=6.15cm]{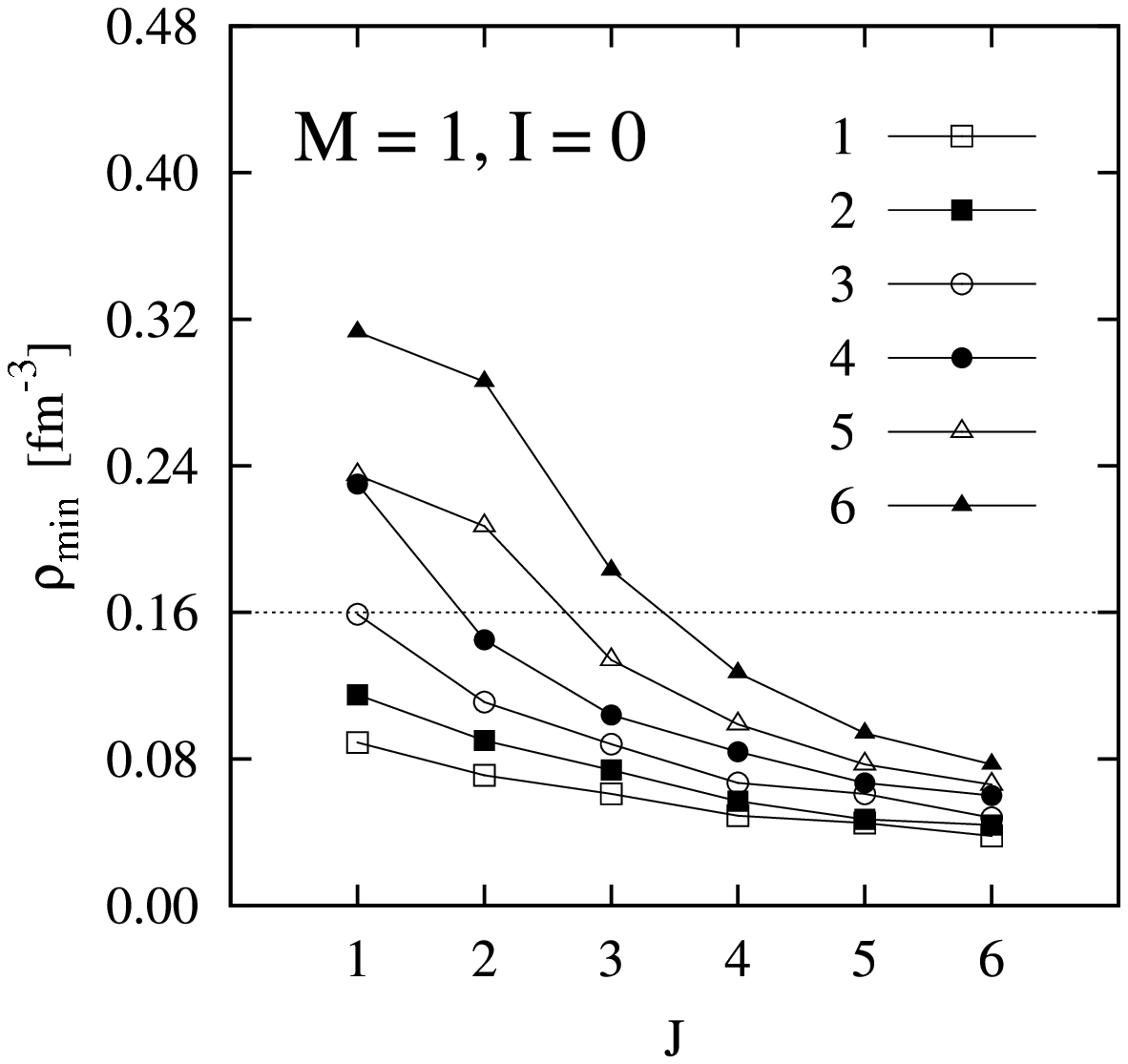}
                 \,&\, \includegraphics[angle=270,width=6.15cm]{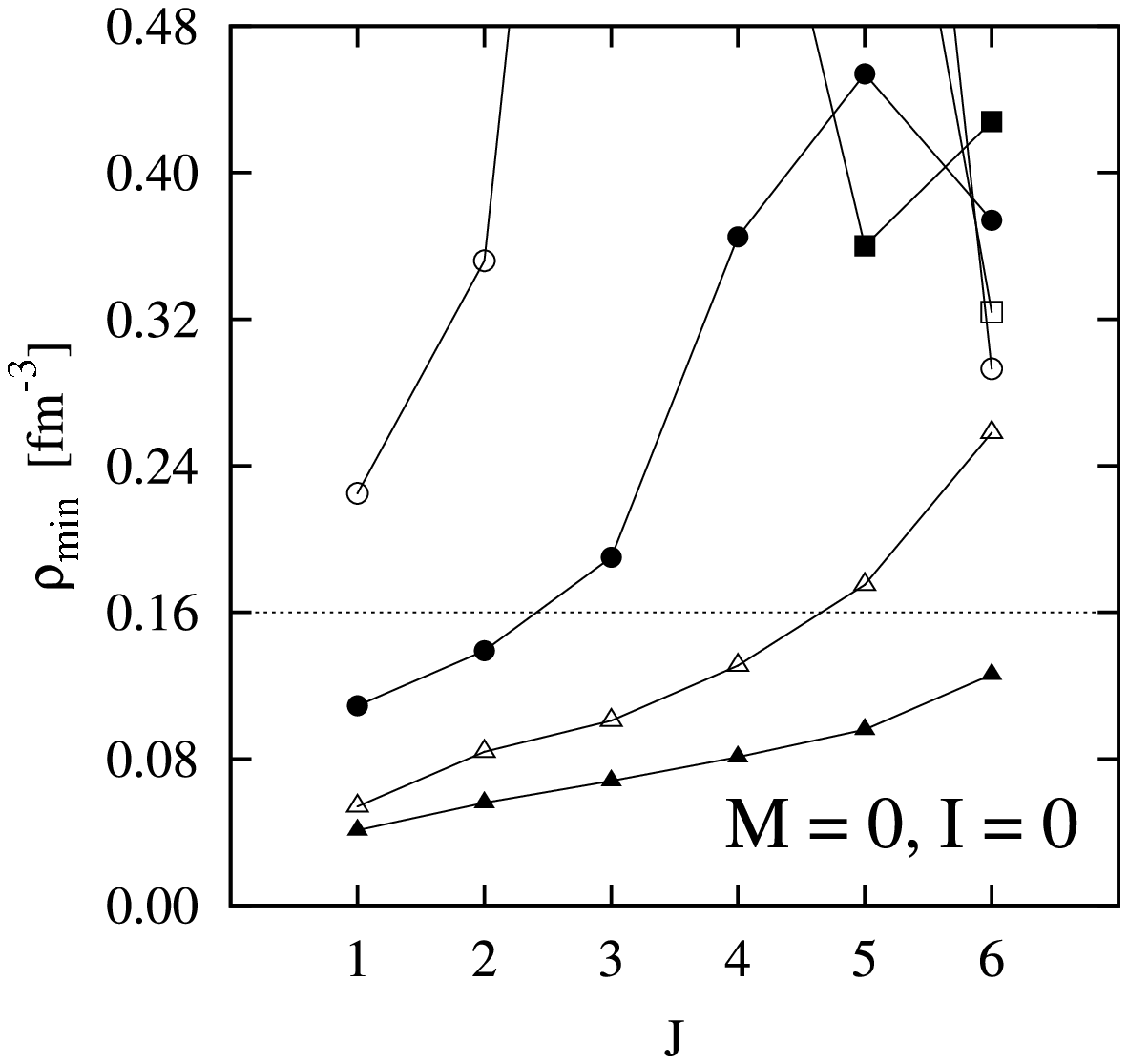} \\
\includegraphics[angle=270,width=6.15cm]{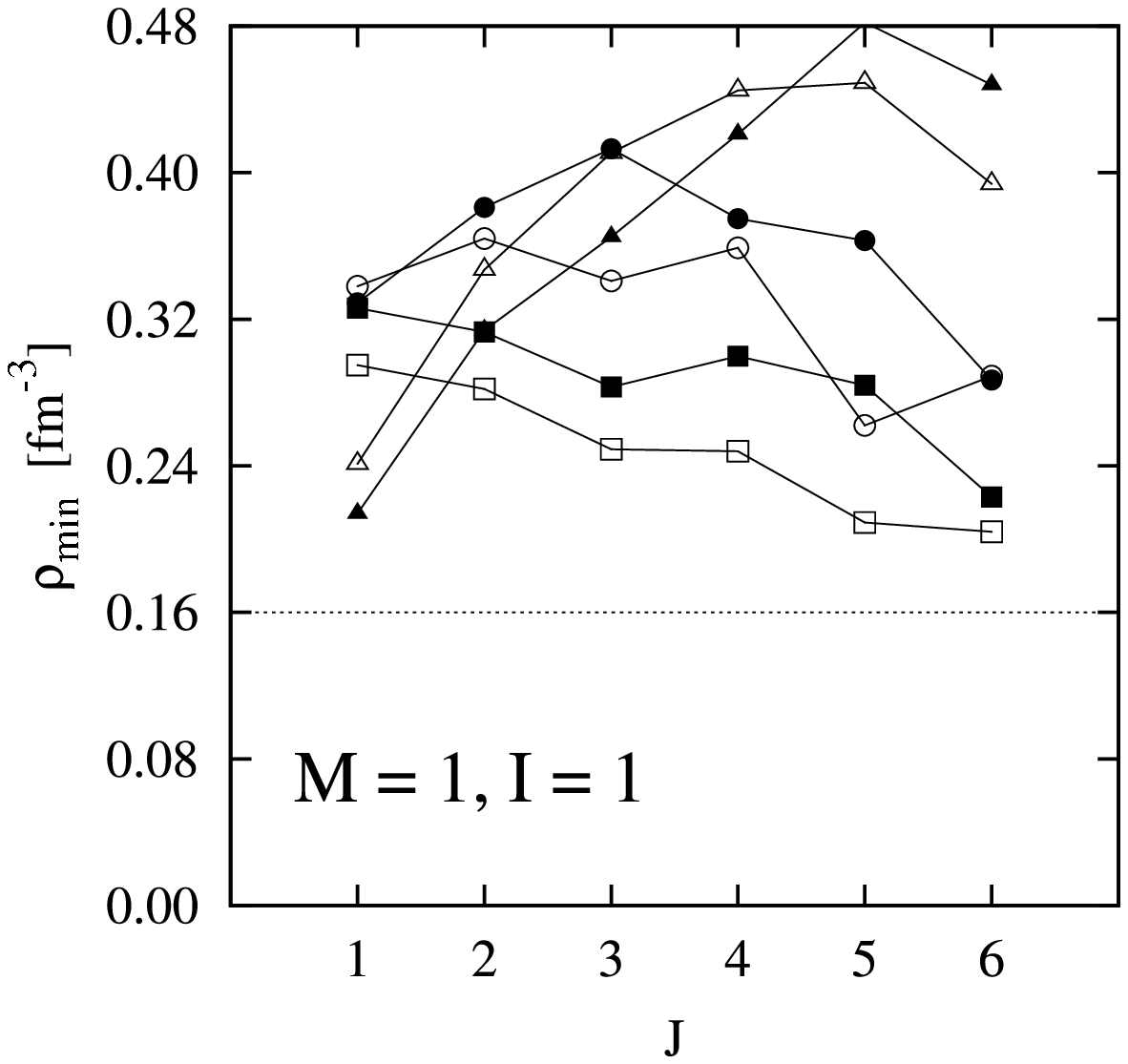}
                 \,&\, \includegraphics[angle=270,width=6.15cm]{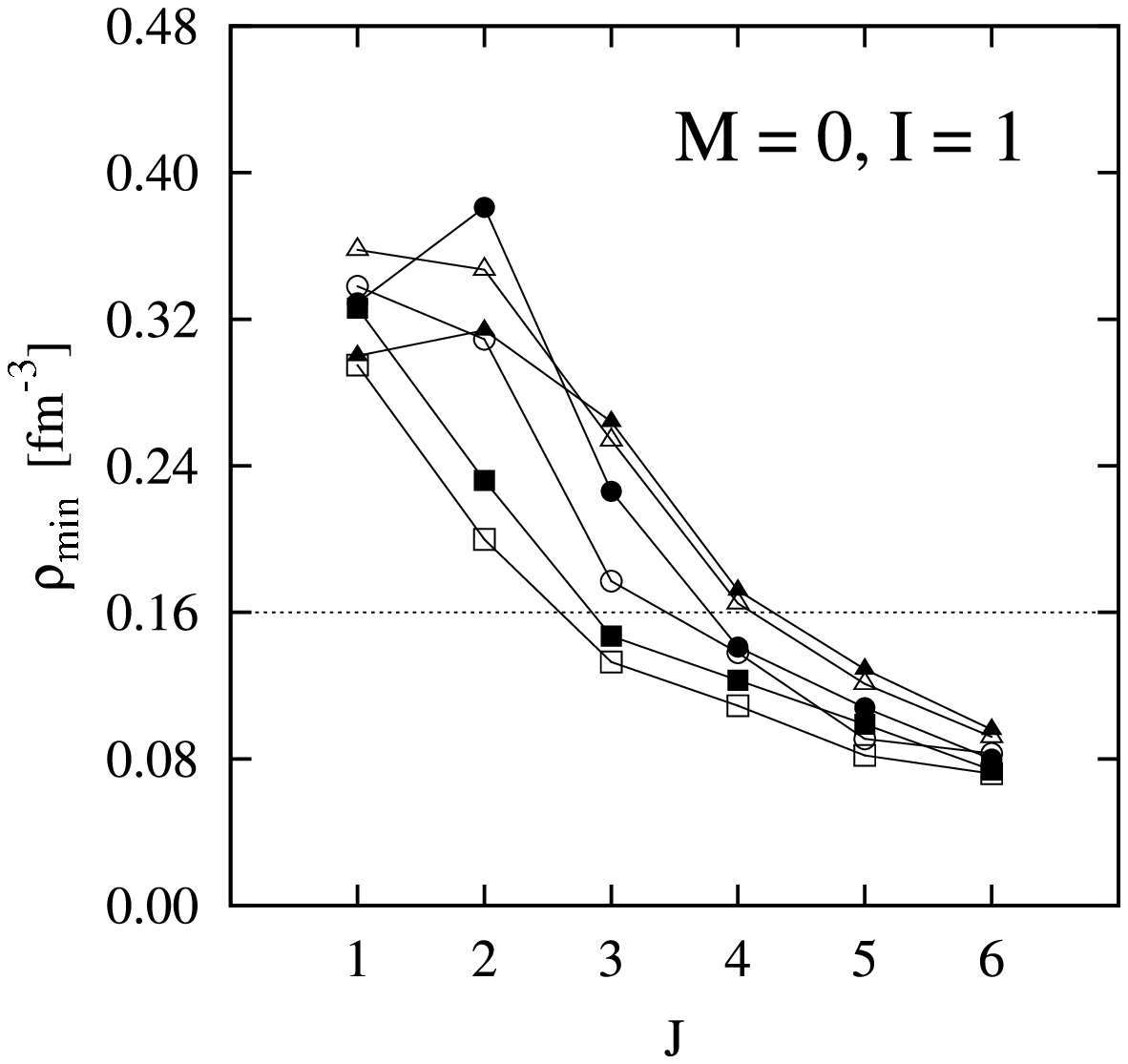} \\
\end{matrix}$
\vspace*{8pt}
\caption{Minimal value of the critical densities in the four channels
with $S=1$ for the T$IJ$ interactions. The correspondance between
the symbols and the values of $I$ is given on the top left panel. The $J$
index of the interactions corresponds to the coordinate on the horizontal
axis. The horizontal dotted lines represent the saturation density of the
system $\rho_\mathrm{sat}$.
\label{fig:tij}}
\end{figure}

The instabilities in the spin channels can not be shown as easily
as the isospin ones since one has to break the time reversal symmetry
and therefore the spherical symmetry as well.
In Fig.~\ref{fig:tij} we show the minimal value $\rho_\mathrm{min}$
of the critical densities $\rho_\mathrm{crit}$ for any $q$ for all the
T$IJ$ interactions in the
channels with $S=1$. To have a stable interaction, it is necessary that the
functional satisfies the minimal condition $\rho_\mathrm{min}>
\rho_\mathrm{sat}$ in all channels. As a result we observe that none of
the T$IJ$ interactions is stable, this is in agreement with recent
cranked HFB calculations in finite nuclei\cite{hellemans}.

\section{Conclusion}

We have given and illustrated an example of the finite size spin and isospin
instabilities that can occur in nuclei. We have also reported that
finite size instabilities are known to appear in calculations
that break the time reversal symmetries with several Skyrme interactions.

The terms of the functional that are responsible for
the appearance of the instabilities can be indentified and omitted,
but this rough method breaks the link between the functional and
an underlying effective interaction. Calculations beyond the mean-field
approximation are then more involved since one has to remove spurious
contributions from the energy. If one wants to keep the link that
relates the interaction to the functional and avoid those spurious
contributions in the energy, one needs a method that can predict
there appearance and this method must be numerically fast enough
to be usable during the fitting procedure of the parameters.

The method of the linear reponse that we present here seems to be
an efficient tool to reveal these finite size instabilities, although
the exact correspondance between the instabilities in infinite
nuclear matter and nuclei needs to be studied in more details.
First of all, finite size and shell effects being non trivial
in nuclei, one has to test whether there is a robust correspondance
between instabilities in infinite nuclear matter and nuclei.
If this is confirmed, one has to estimate how close the critical
densities can be from the saturation density before an instability
shows up.

The calculations of the linear response and critical densities
being very fast, they will be incorporated in the
fitting procedure of effective forces and functionals in the future.

\section*{Acknowledgements}

This work was supported by the NESQ project (ANR-BLANC 0407).


\begin{thebibliography}{0}

\bibitem{gogny75}
D. Gogny, {\em Nucl. Phys. A} {\bf 237} (1975) 399.

\bibitem{gogny80}
J. Decharg\'e, D. Gogny, {\em Phys. Rev. C} {\bf 21} (1980) 1568.

\bibitem{skyrme56}
T.H.R. Skyrme, {\em Phil. Mag.} {\bf 1} (1956) 1043;
J.S. Bell and T.H.R. Skyrme, {\em Phil. Mag.} {\bf 1} (1956) 1055.

\bibitem{skyrme58}
T.H.R. Skyrme, {\em Nucl. Phys.} {\bf 9} (1958) 615;
 {\em Nucl. Phys.} {\bf 9} (1958) 635.

\bibitem{lacroix09}
D. Lacroix, T. Duguet, M. Bender,
{\em Phys. Rev. C} {\bf 79} (2009) 044318.

\bibitem{bender09}
M. Bender, T. Duguet, D. Lacroix,
{\em Phys. Rev. C} {\bf 79} (2009) 044319.

\bibitem{duguet09}
T. Duguet, M. Bender, K. Bennaceur, D. Lacroix, T. Lesinski,
{\em Phys. Rev. C} {\bf 79} (2009) 044320.

\bibitem{washiyama}
K. Washiyama, K. Bennaceur, J. Meyer, M. Bender, P.-H. Heenen and
V. Hellemans, {\em in~prep}.

\bibitem{sadoudi}
J. Sadoudi, thesis, 2011.

\bibitem{chabanat} E. Chabanat, P. Bonche, P. Haensel, J. Meyer and
R. Schaeffer,
{\it Nucl. Phys. A} {\bf 627} (1997) 710;
{\it Nucl. Phys. A} {\bf 635} (1998) 231;
{\it Nucl. Phys. A} {\bf 643} (1998) 441(E).

\bibitem{lesinski07}
T. Lesinski, M. Bender, K. Bennaceur, T. Duguet and J. Meyer,
{\it Phys. Rev. C} {\bf 76} (2007) 014312.

\bibitem{bender09b}
M. Bender, K. Bennaceur, T. Duguet, P.-H. Heenen, T. Lesinski and J. Meyer,
{\em Phys. Rev. C} {\bf 80} (2009) 064302.

\bibitem{lesinski06} T. Lesinski, K. Bennaceur, T. Duguet and J. Meyer,
{\it Phys. Rev. C} {\bf 74} (2006) 044315.

\bibitem{colo10}
L.~Cao, G.~Col\`o, H.~Sagawa,
{\it Phys. Rev. C} {\bf 81} (2010) 044302.

\bibitem{hellemans}
V.~Hellemans, P.-H.~Hennen, private communication.

\bibitem{schunck}
N.~Schunck, K.~Bennaceur, D.~Davesne, T.~Duguet, T.~Lesinski and
A.~Pastore, {\em in~prep}.

\bibitem{garciarecio}C. Garc\'ia-Recio, J. Navarro, N. Van Giai and
L.L. Salcedo,
{\em Ann. Phys. (NY)} {\bf 214} (1992) 293.

\bibitem{fetter}A. L. Fetter and J. D. Walecka,
\emph{Quantum Theory of Many-Particle Systems} (McGraw-Hill, New-York, 1971).

\bibitem{davesne} D. Davesne, M. Martini, K. Bennaceur and J. Meyer,
{\it Phys. Rev. C} {\bf 80} (2009) 024314;
{\it Phys. Rev. C} {\bf -} (2011) - (E).

\end{thebibliography}
\end{document}